\documentclass[journal]{IEEEtran}
\usepackage[utf8]{inputenc}
\usepackage[english]{babel}
\usepackage{amsthm}
\usepackage{amsfonts,amssymb,amsthm}
\usepackage{bbm}
\usepackage{mathtools}
\usepackage{physics}
\usepackage{amsfonts}
\usepackage{enumerate}
\usepackage{mathtools}
\usepackage{stackengine}

\usepackage{mathrsfs}
\usepackage{subcaption}

\newtheorem{theorem}{Theorem}
\newtheorem{lemma}{Lemma}
\newtheorem{corollary}{Corollary}
\newtheorem{remark}{Remark}
\newtheorem{proposition}{Proposition}
\newtheorem{definition}{Definition}

\usepackage{pgfplots}
\usepackage{array,colortbl,multirow,multicol,booktabs,ctable}

\usepackage{floatrow}
\usepackage{graphicx}
\DeclareMathOperator*{\argmax}{arg\,max}

\usepackage{algorithm}
\usepackage{algpseudocode}
\usepackage{enumitem}

\pgfplotsset{compat=1.18} 

\title{{Ergodic Capacity and Optimal Handover in Satellite Mega-Constellations under Finite Serving Times}}

\author{Brendon~McBain,~\IEEEmembership{Member,~IEEE,}
        Yi~Hong,~\IEEEmembership{Senior Member,~IEEE,}
        and~Emanuele~Viterbo,~\IEEEmembership{Fellow,~IEEE}
\thanks{A preliminary version of part of this work appeared in \cite{Mcbain2025b}.}
\thanks{This research work was supported by the Australian Research Council (ARC) through the Discovery Project under Grant DP260101376.}%
\thanks{The authors are with the Department of Electrical and Computer
Systems Engineering, Monash University, Clayton, VIC 3800, Australia
(e-mail: \{brendon.mcbain, yi.hong, emanuele.viterbo\}@monash.edu).}%
\thanks{Corresponding authors: Brendon Mcbain and Emanuele Viterbo.}
}

\begin{document}

\maketitle

\begin{abstract}
    Existing analyses of ergodic capacity in satellite mega-constellations often rely on restrictive serving time assumptions or become intractable under realistic handover strategies. This paper develops a framework for characterising the ergodic capacity of low-Earth-orbit (LEO) mega-constellation links under arbitrary handover strategies and serving times. The user--satellite link is modelled as shadowed-Rician fading, and a semi-stochastic satellite channel with persistence is introduced in which visible satellites are drawn from a non-homogeneous binomial point process (NBPP) at each handover and the selected satellite is then propagated using circular orbit dynamics. Under uncoordinated handover decisions, this yields independent serving periods and enables a renewal-theoretic derivation of persistent capacity. This capacity is related to the non-persistent capacity from prior work, and closed-form bounds are provided for efficient evaluation. Optimal handover is then formulated as a non-linear fractional program, yielding an explicit decision rule via a variant of Dinkelbach's algorithm. The results show that a simpler strategy that maximises serving capacity closely approximates the optimum while performing best under SGP4-based orbit prediction and mega-constellation simulation.
\end{abstract}

\section{Introduction}

The dynamic architecture of {\em low-Earth orbit (LEO)} satellite networks presents unique challenges in maintaining continuous and reliable communication, particularly during satellite handovers \cite{LiuXiao2024,Akyildiz1999,LiuYang2024,Chowdhury2006}. The rapid movement of satellites in the mega-constellation necessitates efficient handover strategies that minimise service degradation for communications between LEO satellite networks and ground users. In this context, information-theoretic handover strategies are critical for achieving high-throughput communications in the presence of frequent transitions between serving satellites. 
{This paper develops a tractable semi-stochastic framework for analysing and optimising \emph{user-centric} handovers in the ergodic capacity sense \emph{under channel persistence}.}

The communications system model is often based on a time-varying satellite constellation model with satellites serving multiple users within their spotbeams on Earth. This deterministic system model is used in simulation-based analyses for an accurate description of the performance of a particular system. However, it poses tractability issues for theoretical analysis, especially for ultra-dense LEO mega-constellations that can comprise hundreds or thousands of satellites \cite{DelPortillo2019}. Taking advantage of the large number of satellites in an ultra-dense mega-constellation, stochastic system models have been developed in recent years to model satellite mega-constellations using point processes \cite{Okati2020a, Okati2022}. While stochastic models are limited in their ability to accurately capture time-dependent orbit trajectories over short time scales, they offer tractable models that are amenable to general theoretical analyses. The loss in accuracy of stochastic models can be reduced using our {\em semi-stochastic} model \cite{Mcbain2025b}, which recovers the time-dependent behaviour of orbit trajectories while retaining tractability. {Specifically, we retain deterministic trajectories only for the satellites visible at the handover decision time, which preserves the relevant time dependence for handover optimisation while keeping the analysis tractable.} 
{More broadly, whether a deterministic, stochastic, or semi-stochastic model is appropriate depends on the purpose of the analysis; for handover optimisation, incorporating time dependence is essential whenever satellites persist long enough for link statistics to evolve.}

A critical aspect of satellite mega-constellation networks is the {\em handover strategy}{, which} determines which satellite from the mega-constellation {serves} the ground user and for how long. {In} ultra-dense mega-constellations, there are typically tens of visible satellites that are candidates for serving the user. {By exploiting} this satellite diversity, the system has the unique ability to {select} the user--satellite channel through the handover decision. In addition, {each} handover decision is {initiated by} a {\em handover trigger}, {namely} the event that {prompts} evaluation of the candidate satellites at the current moment. Therefore, the handover strategy---which specifies both the {\em decision rule} and the trigger---is critical to system performance.

For example, the handover decision rule may be to choose the satellite with the minimum distance{, or equivalently the maximum elevation angle,} since {this} minimises path loss and propagation delay. {The} handover trigger may {then} be the event that the serving satellite is no longer visible{, that is, falls} below the minimum elevation angle, since the lack of a line-of-sight {link} significantly degrades the channel. In \cite{Voicu2024}, Voicu \emph{et al.} conducted an empirical analysis of this strategy in terms of capacity, propagation delay, and Doppler shift. Motivated by this example, handover strategies are {often} designed {according to} Papapetrou's criteria \cite{Papapetrou2004}:
\begin{enumerate}
    \item {\em Maximum serving time}: Choose the satellite with the longest serving time to avoid frequent satellite handover;
    \item {\em Minimum distance}: Choose the nearest satellite, as it provides the lowest path {loss} and propagation delay;
    \item {\em Maximum number of free channels}: Choose the satellite with the largest number of free channels to balance the network load.
\end{enumerate}
In {the preceding} example, the decision rule was based on Criterion~2 and the trigger was based on Criterion~1. However, choosing the nearest satellite as in Criterion~2 does not always yield the most reliable user--satellite channel in the presence of shadowing. In addition, since this criterion is {applied only} at the handover time, it does not capture the time-dependent evolution of the user--satellite channel {before the next} trigger. If serving times are long enough for substantial variations in reliability, such a handover decision rule is {therefore} unlikely to be optimal in aggregate. {Criterion~3, on the other hand,} necessarily involves a multi-user scenario with limited resources, and handover strategies that explicitly {incorporate} this criterion often lose tractability due to dependencies between users \cite{Wu2016}. While this does not directly degrade physical-layer link quality, it does limit the ability to evaluate and optimise handover strategies. 
{ These observations motivate an objective that explicitly accounts for persistence, leading to the persistent-capacity formulation adopted in this work.}

{
\subsection{Background and Related Work}

Existing analytical studies of LEO mega-constellation handover have mainly considered either stochastic constellation models \cite{Okati2020a,Okati2022} without explicit time dependence or handover rules based on instantaneous link quality \cite{Voicu2024}. For example, \cite{Okati2022b} formalised an \emph{optimal} handover strategy that maximises received SNR and hence ergodic capacity, while \cite{Guo2024} proposed related strategies for multi-tier mega-constellations. Such works primarily address Papapetrou's Criterion~2, whereas Criterion~1 requires modelling persistence whenever serving times are long enough for link quality to evolve over time. As shown later through a special case of our framework, purely stochastic models are therefore most appropriate for short serving times.

Papapetrou's Criterion~3 is fundamentally multi-user, since free-channel availability and load balancing must be handled through network-level coordination. This makes fully integrated multi-user handover optimisation difficult to analyse tractably. Our focus in this paper is therefore user-centric: we optimise the long-term physical-layer utility associated with an individual user's serving process. Nevertheless, the resulting methods remain relevant in multi-user networks because they provide, for each user, a principled utility metric and corresponding ranking over candidate satellites based on persistent link quality. Such user-specific rankings can be passed to higher-layer assignment or scheduling algorithms \cite{Wu2016} to enforce load, interference, and resource constraints. In this sense, the present framework does not solve the multi-user allocation problem itself, but provides a tractable physical-layer basis that can reduce the search space and improve the quality of candidate selections considered by network-level coordination algorithms.}

{ Motivated by these limitations, we adopt a \emph{semi-stochastic} model that supports arbitrary serving times through the notion of \emph{persistent (satellite) capacity} \cite{Mcbain2025b}. In this setting, the persistent-capacity-optimal user-centric handover decision requires orbit prediction to evaluate the serving capacity of candidate satellites over their prospective serving intervals. The resulting framework captures persistence while remaining analytically tractable, thereby bridging purely stochastic models and fully deterministic orbit-based simulations.}

\subsection{Contributions}

{ The main contributions are as follows:
\begin{itemize}
    \item We develop a semi-stochastic framework for user-centric handover in LEO satellite networks, yielding a renewal channel model termed the \emph{persistent satellite channel}. { Relative to \cite{Mcbain2025b}, the framework additionally includes closed-form orbit prediction equations and an explicit NBPP sampling method for efficient computation and simulation of time-dependent performance metrics.}

    \item We derive the ergodic capacity of the persistent satellite channel, termed the \emph{persistent capacity}, relate it to the non-persistent capacity in prior work \cite{Okati2020b}, and obtain closed-form upper and lower bounds. { This extends \cite{Mcbain2025b} by clarifying the relation between persistent and non-persistent formulations and by providing new analytical bounds.}

    \item We formulate optimal handover as the maximisation of persistent capacity and solve it via a variant of Dinkelbach's algorithm, yielding a closed-form decision rule. { We further show that the maximum serving capacity handover strategy from \cite{Mcbain2025b} closely approximates the optimal rule.}

    \item { We validate the proposed framework through SGP4-based orbit prediction and simulation of the Starlink mega-constellation, including both unconstrained serving times and fixed serving times of $15$ seconds. The results show that the maximum serving capacity handover strategy performs best in this more realistic setting, making it a practical low-complexity approximation to optimal handover.}
\end{itemize}

The remainder of this paper is organised as follows. Section \ref{sec:modelling} introduces the semi-stochastic mega-constellation model. Section \ref{sec:channel} defines the handover strategy and the persistent satellite channel model. Section \ref{sec:capacity} derives the persistent capacity. Section \ref{sec:handover} develops the theoretical analysis of optimal handover strategies. Section \ref{sec:numerical_results} presents numerical results on ergodic capacity and handover.}

\subsection{Notation} A random variable is denoted by a capital letter such as $X$. The probability density of $X$ is denoted by $f_X(x)$ for $x\in\mathbb{R}$. The expectation of $g(X)$, for some function $g$, is $\mathbb{E}[g(X)] = \int_{\mathbb{R}} g(x) f_X(x) dx$. For functions of multivariate random variables, e.g., $X$ and $Y$, we denote the expectation with respect to $X$ as $\mathbb{E}_X[g(X,Y)]$ where $Y$ is being conditioned on. The i.i.d. continuous uniform random variable is denoted by $\mathcal{U}(\mathcal{A})$ with compact support $\mathcal{A}\subset \mathbb{R}$. The base-$2$ logarithm is denoted by $\log$.

\section{Modelling Satellite Mega-Constellations}\label{sec:modelling}

\begin{figure*}
    \centering
    \includegraphics[trim={12cm 3cm 9.5cm 4.9cm},clip,width=0.70\textwidth]{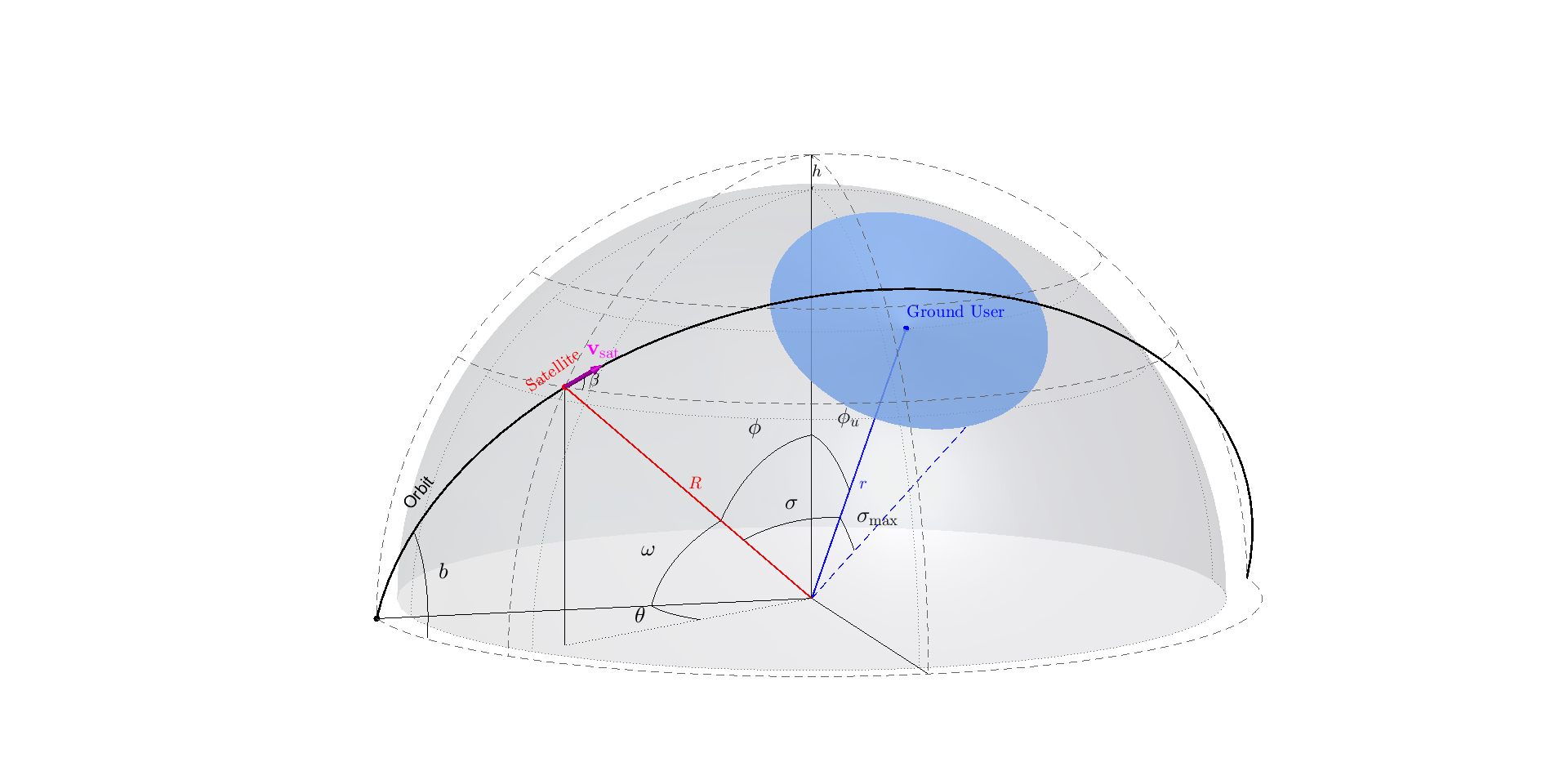}
    \caption{A satellite on an orbit towards the visibility cap of a ground user.}
    \label{fig:LEOorbit}
\end{figure*}

\subsection{Visibility Cap of a Ground User}

In practice, the set of satellites that can serve a user may be reduced by local blockage and/or interference-mitigation constraints. It is therefore common to impose a minimum elevation angle $\psi_{\min}$ above the user horizon, and to declare a satellite \emph{visible} only if its elevation exceeds $\psi_{\min}$. Equivalently, visibility corresponds to a spherical cap on the satellite sphere {of radius $R=r+h$}, bounded by a maximum user--satellite central angle $\sigma_{\max}$ determined by $\psi_{\min}$, {the Earth radius $r$, and the satellite altitude $h$}.

Consider a user at $(r,\theta_{\rm u},\phi_{\rm u})$ and a satellite at $(R,\theta,\phi)$, as illustrated in Fig. \ref{fig:LEOorbit}. {The central angle between their position vectors is denoted by $\sigma(\theta,\phi; \theta_{\rm u},\phi_{\rm u})$, which we often write as $\sigma(\theta,\phi)$ for a fixed user location,} and is given by
\begin{equation}
\label{eq:central_angle}
\begin{split}
\sigma(\theta,\phi)
&=
\cos^{-1}\!\big(
\cos\phi_{\rm u}\cos\phi \\
&\qquad\qquad
+\sin\phi_{\rm u}\sin\phi\cos(\theta_{\rm u}-\theta)
\big).
\end{split}
\end{equation}
The visible region is the spherical cap
\begin{equation}
\label{eq:vis_cap}
\mathsf{Cap}
=
\big\{(\theta,\phi) \in [0,2\pi) \times [0,\pi): \sigma(\theta,\phi)\le \sigma_{\max}\big\}.
\end{equation}

For integration over $(\theta,\phi)$ it is convenient to parameterise the boundary $\sigma(\theta,\phi)=\sigma_{\max}$ by solving for $\theta$ as a function of $\phi$. {The cap intersects the polar angle range $\phi\in[\phi_{\min},\phi_{\max}]$,
\begin{align}
\label{eq:visible_phi_support}
\phi_{\min}&=\max\{0,\phi_{\rm u}-\sigma_{\max}\},\\
\qquad
\phi_{\max}&=\min\{\pi,\phi_{\rm u}+\sigma_{\max}\}.
\end{align}}
For $\phi\in[\phi_{\min},\phi_{\max}]$, solving $\sigma(\theta,\phi)=\sigma_{\max}$ yields \cite{Mcbain2025a}
\begin{equation}
\theta_L(\phi)=\theta_{\rm u}-\tfrac{1}{2}L(\phi),
\qquad
\theta_U(\phi)=\theta_{\rm u}+\tfrac{1}{2}L(\phi),
\end{equation}
where $L(\phi)$ is the \emph{longitude span} of the cap at polar angle $\phi$:
\begin{align}
\label{eq:longitude_span}
L(\phi)
&=
\pi
+
2\sin^{-1}\!\Big(
\csc\phi\Big(\cot\phi_{\rm u}\cos\phi \notag\\
&\qquad\qquad\qquad\qquad\qquad-\cos\sigma_{\max}\,\csc\phi_{\rm u}\Big)
\Big).
\end{align}
Hence,
\begin{equation}
\label{eq:cap_param}
\mathsf{Cap}
=
\big\{(\theta,\phi):\
\theta_L(\phi)\le \theta \le \theta_U(\phi),\
\phi\in[\phi_{\min},\phi_{\max}]
\big\},
\end{equation}
which avoids the implicit constraint $\sigma(\theta,\phi)\le \sigma_{\max}$ and is convenient for subsequent surface integrals.

\subsection{Deterministic Model}
\paragraph{Circular-orbit approximation}
Consider a LEO orbit at altitude $h$ around a static and spherical Earth of radius $r$. A satellite is moving along this orbit on a satellite sphere of radius $R=r+h$ as illustrated in Fig. \ref{fig:LEOorbit}. { For a {\em prograde orbit}, the {\em argument of latitude} $\omega$ is the angle in the orbital plane measured from the {\em ascending node} to the satellite position, with $\omega=0$ at the ascending node, where the satellite crosses the Earth's equatorial plane from south to north. The orbital plane is inclined to the equatorial plane by the inclination angle $b$. The satellite is ascending in latitude when $|\omega|<\pi/2 \pmod{2\pi}$, and descending in latitude otherwise.} The satellite is moving at a constant speed $v_{\rm{sat}}$, resulting in angular velocity $\omega_{\rm{sat}} = v_{\rm{sat}}/R$ and orbital period $T_{\rm{sat}} = 2\pi/\omega_{\rm{sat}}$. The direction of the velocity vector of the satellite and the latitude line at $\phi$ is $\beta_a(\phi) = a\cos^{-1}(\cos(b)/\sin(\phi))$ where $a={+1}$ if ascending or $a={-1}$ if descending. Satellite mega-constellations are composed of multiple orbital planes with ascending nodes spaced by angle $s_{\rm{orb}}$, and each orbital plane contains $N_{\rm{orb}}$ uniformly distributed satellites. The total number of satellites is $N_{\rm{sat}} =  2\pi N_{\rm{orb}}/ s_{\rm{orb}}$. 

{
{Consider a satellite initialised at position $(R,\theta(0),\phi(0))$ in direction $a$ on a circular orbit. The position $t$ seconds in the future can be} computed using a sequence of rotations: (1) Initialise the satellite on a flat orbital plane at spherical coordinates $(R, \omega_{\rm{sat}}t, 0)$;
    (2) rotate counter-clockwise around the x-axis by satellite direction $\beta_a(\phi(0))$ radians;
    (3) rotate clockwise around the y-axis by latitude $\frac{\pi}{2} - \phi(0)$ radians; 
    (4) rotate counter-clockwise around the z-axis by longitude $\theta(0)$ radians. { This yields the following theorem, which gives closed-form orbit prediction equations used later to explicitly compute time-dependent performance metrics.}

    \begin{theorem}\label{thm:circ_orbit}
Consider a satellite initialised at polar angle $\phi(0)$ and longitude $\theta(0)$, moving in direction $a$. Let $\psi(t) = \omega_{\rm sat} t$ and $\beta(0)
= \beta_{a}(\phi(0))$, $a\in\{+1,-1\}$.
Define the auxiliary functions
\begin{align}
A(t)
&=
\sin\phi(0) \cos\psi(t)
-
\cos\phi(0) \sin\beta(0)\, \sin\psi(t),
\\
B(t)
&=
\cos\beta(0)\, \sin\psi(t),\\
C(t) &= \cos\phi(0) \cos\psi(t) + \sin\phi(0) \sin\beta(0) \sin\psi(t).
\end{align}
At time $t$, the satellite has polar angle $\phi(t)$ and longitude $\theta(t)$, given by
\begin{align}
\phi(t)
&=
\cos^{-1}(
C(t)),
\\
\theta(t)
&=
\theta(0)
+
\operatorname{atan2}\!\big(
B(t),\, A(t)
\big).
\end{align}
\end{theorem}
\begin{proof}
    The proof is given in the Appendix.
\end{proof}
}

Hence, the time-dependent distance between a user and the satellite is $d(t)=d(\theta(t), \phi(t))$ and their time-dependent central angle is $\sigma(t)=\sigma(\theta(t),\phi(t))$ using Theorem \ref{thm:circ_orbit}. In addition, the satellite only has a line-of-sight with the user for a visibility time $T_{\rm vis}(\theta(0), \phi(0),a)$, in seconds, and is the time it takes for the elevation angle of the satellite to equal the minimum elevation angle $\psi_{\min}$ (or maximum central angle $\sigma_{\max}$). Formally, this can be found by finding the root of $\sigma(t) - \sigma_{\max}$ over domain $t \in [0,T_{\rm{sat}}/2]$, noting that this domain excludes the root on the other side of the visibility cap of satellites.

\paragraph{SGP4 orbit propagation}
Realistic orbit trajectories are generated using the \emph{Simplified General Perturbations 4 (SGP4)} propagator, which accounts for key perturbations (e.g., Earth's oblateness, atmospheric drag, solar radiation pressure, and third-body gravity from the Sun and Moon) that cause departures from ideal circular motion. Given a constellation \emph{two-line element (TLE)} set, SGP4 produces the corresponding time-varying satellite ephemerides.

\subsection{Stochastic Model}
We now leverage the fact that there are many satellites in a LEO mega-constellation, so that we can stochastically model the satellite positions using a point process. The uniformly spaced ascending nodes are stochastically modelled as a continuous uniform variable, such that the rotational angle (longitude) of a satellite  is a random variable
\begin{align}
    \Theta \sim \mathcal{U}[0,2\pi) .
\end{align}

 The uniformly spaced satellites on an orbital plane are stochastically modelled as a continuous uniform random variable for the argument of latitude, resulting in a polar angle $\Phi$ with PDF (as derived in \cite[Lemma~2]{Okati2020b} for latitude $\pi/2 - \phi$)
        \begin{align}
    f_{\Phi}(\phi) &= \frac{\sin(\phi)}{\pi \sqrt{\sin^2(\phi)-\cos^2(b)}} 
\end{align}
for $\phi \in [\pi/2-b,\pi/2+b]$ and zero otherwise. Thus, the positions of the $N_{\rm sat}$ satellites in the mega-constellation are i.i.d. stochastically modelled as the spherical coordinates $(R,\Theta,\Phi)$, forming an NBPP on a sphere \cite{Mcbain2025a}. The NBPP is an accurate model for the first-order statistics of the positions of satellites in a mega-constellation, assuming the orbital spacing $s$ is small and the satellites spend equal durations at all longitudes and all arguments of latitude. In addition, we mark the satellite positions in the NBPP with a uniform binary variable $A$ that specifies whether the satellite is ascending (${+1}$) or descending (${-1}$) on its orbit, which is the sign of direction angle $\beta_A(\Phi)$. 

Since we are often interested in the influence of the mega-constellation on a ground user, we can form a user-centric stochastic model by defining the random set of $N_{\rm vis}$ visible satellites as {$\mathcal{V}=\{(\Theta_1,\Phi_1,A_1),(\Theta_2,\Phi_2,A_2),\cdots\}$} with respect to a user at $(r,\theta_{\rm u},\phi_{\rm u})$. Observe that in the stochastic model there is a non-zero probability that we have the event $N_{\rm vis}=0$, which would not actually be possible in a real mega-constellation designed for full ground coverage; hence, we include the condition $N_{\rm vis} \geq 1$ in the model.

{The mega-constellation NBPP model can be sampled using uniform random variables as follows. This is important for efficient simulation of performance metrics that are otherwise intractable and do not admit closed-form expressions.}

\begin{theorem}\label{thm:sampling_thm}
Conditioned on having exactly $N$ visible satellites, the visible satellite
locations are sampled independently as follows. First generate a proposal
polar angle
\begin{equation}
    \Phi^\star
    =
    \cos^{-1}\!\left(\sin b\sin U\right),
    \qquad
    U\sim\mathcal{U}[B,A],
\end{equation}
where
\begin{equation}
    A=\sin^{-1}\!\left(\frac{\cos\phi_{\min}}{\sin b}\right),
    \qquad
    B=\sin^{-1}\!\left(\frac{\cos\phi_{\max}}{\sin b}\right).
\end{equation}
{Accept $\Phi^\star$ with probability
\begin{equation}
    \frac{L(\Phi^\star)}{L_{\max}},
    \qquad
    L_{\max}
    =
    \max_{\phi\in[\phi_{\min},\phi_{\max}]}L(\phi).
\end{equation}} 
If accepted set $\Phi=\Phi^\star$ otherwise repeat. Conditioned on $\Phi$,
the longitude is generated as
\begin{equation}
    \Theta\mid\Phi
    \sim
    \mathcal{U}[\theta_L(\Phi),\theta_U(\Phi)].
\end{equation}
Repeating until $N$ samples are accepted gives exactly $N$ visible satellites
from the conditional NBPP model.
\end{theorem}

{
\begin{proof}
The proposal $\Phi^\star$ is obtained by restricting the orbital polar angle
law to $[\phi_{\min},\phi_{\max}]$, and hence has density proportional to
$f_\Phi(\phi)$ on this interval. Conditioning on visibility further weights
each polar angle by the visible longitude span $L(\phi)$, since only a
longitude interval of length $L(\phi)$ is visible at polar angle $\phi$.
Thus the desired visible polar angle density is proportional to
$f_\Phi(\phi)L(\phi)$.

The acceptance probability $L(\Phi^\star)/L_{\max}$ converts the proposal
density into this desired density, because the accepted samples have density
proportional to
\[
    f_\Phi(\phi)\frac{L(\phi)}{L_{\max}}
    \propto
    f_\Phi(\phi)L(\phi).
\]
Conditioned on the accepted polar angle, visibility only restricts the
longitude to $[\theta_L(\Phi),\theta_U(\Phi)]$, so the conditional longitude is
uniform on this interval. Repeating the independent sampling procedure until
$N$ samples are accepted gives exactly $N$ visible satellites.
\end{proof}}

{
Theorem~\ref{thm:sampling_thm} provides an efficient conditional sampler for generating exactly $N$ visible satellites from the NBPP model. Rather than sampling satellites over the entire satellite sphere and rejecting those outside the visibility cap, the sampler first restricts the polar angle to the range that can intersect the cap. It then accounts for the remaining visibility constraint through the longitude span $L(\phi)$ available at each polar angle. This avoids wasting samples on satellites that cannot be visible and is useful for Monte Carlo simulation of user-centric performance metrics.}

\section{Persistent Satellite Channel}\label{sec:channel}
The information-bearing signal \( x(t) \) is transmitted by a serving satellite \( (\Theta_0, \Phi_0, A_0) \in \mathcal{V} \) with orbit trajectory \( (R, \Theta_0(t), \Phi_0(t)) \) and is received as \( y(t) \) by a fixed ground user at \( (r, \theta_{\rm u}, \phi_{\rm u}) \)\footnote{Without loss of generality, the transmission link can be considered as either the uplink or the downlink.}. Co-channel interference from terrestrial or non-terrestrial networks is assumed mitigated to an acceptable noise level \( N_0 \). The line-of-sight (LOS) signal propagates over a free-space distance \( d(\Theta_0(t), \Phi_0(t)) \), with path loss proportional to \( d^{-2} \), assumed known via ephemeris data.

Shadowing effects arise from atmospheric phenomena (e.g., rain, clouds, ionospheric scintillation) and local obstructions, while scattering induces non-line-of-sight multipaths. Assuming coherent detection, the resulting fading follows a shadowed-Rician model~\cite{Abdi2003}, characterised by the scattering power \( 2b_0 \), Nakagami parameter \( m \), and LOS power \( \Omega \), forming time- and space-varying fading parameters \( S(\theta, \phi, t) = (b_0, m, \Omega) \). These parameters are assumed to be known at the receiver over the serving time, e.g., via weather data or channel estimation under static or slow-moving conditions.

The satellite mega-constellation channel includes two processes:
\begin{enumerate}
    \item \textit{Handover process}: Selects a satellite from \( \mathcal{V} \) based on position, direction, and predicted fading parameters \( S \), and is modelled as a stochastic point process.
    \item \textit{Propagation process}: Models signal propagation between the user and serving satellite as an AWGN channel with time- and space-varying coefficients for path loss and shadowed-Rician fading.
\end{enumerate}

Since these two processes are repeated consecutively, they form a \emph{semi-stochastic continuous-time renewal process}. After discretisation, we have a \emph{discrete-time renewal process} that is used to form a channel model for practical satellite communication systems.

\subsection{Handover Process}
The handover process in satellite mega-constellation networks involves selecting the best available channel from the visible satellites for a specific \emph{serving period}, with decisions made by the {\em central control unit (CCU)}. The CCU uses known satellite position trajectories and fading parameters, denoted by $\{(\Theta_k(t), \Phi_k(t))\}$ and $S$, respectively. However, it cannot predict small-scale multipath fading or the realisations of large-scale shadowing, only their statistical parameters.

To reduce complexity and memory requirements at the receiver, handover decisions are made independently at each handover event. The receiver must also have access to the same channel state information as the CCU to realise the intended performance gains.

The handover trigger is determined by the \emph{serving time}, which is constrained by:
\begin{itemize}
    \item A minimum serving time $T_{\min}$, imposed by practical considerations such as latency spikes and network congestion. If a satellite’s visible time is less than $T_{\min}$, it is considered to go dark once it is no longer visible (though constellations may be designed to avoid this scenario).
    \item A maximum serving time $T_{\max}$, which allows more frequent handovers and better satellite selection, improving quality of service.
\end{itemize}
This is equivalent to computing the serving time by applying a clamping function to the satellite's visible orbit time. Given all of the above, we can formalise the definition of a handover strategy as follows.

\begin{definition}[Handover strategy]\label{def:HO} 
    A handover strategy $\mathsf{H}$ is a function that maps a set of visible satellites $V=\{(\theta_1,\phi_1,a_1),(\theta_2,\phi_2,a_2),\ldots, (\theta_N,\phi_N,a_N)\}$ and shadowing parameters $S$ to a serving satellite as $\mathsf{H}(V,S) = (\theta_k,\phi_k,a_k)\in V$ with serving time $\mathsf{T}(\theta_k,\phi_k,a_k) =  \min\{ \max\{\mathsf{T}_{\rm{vis}}(\theta_k,\phi_k,a_k),T_{\min}\},T_{\max}\}$.
\end{definition}

While we define handover via a serving time trigger, the framework extends to any scalar metric of the satellite state $(V,S)$. For example, one may threshold the achievable rate to enforce a target link rate when feasible \cite{Voicu2024}, inducing a \emph{capacity-guaranteed visibility cap} \cite{McBain2025c}. In contrast, we do not impose per-instant rate constraints; instead, we maximise ergodic capacity over arbitrary serving times.

\subsection{Propagation Process}
The propagation process is initialised with a handover strategy $\mathsf{H}$ applied to the visible satellites $\mathcal{V}$ to get the orbit initialisation $(\Theta_0,\Phi_0,A_0) = \mathsf{H}(\mathcal{V},S)$, which is then used to determine the time-varying channel conditions along the orbit trajectory, while data is transmitted during the serving time.

\begin{figure}
    \centering
    \includegraphics[width=0.99\linewidth]{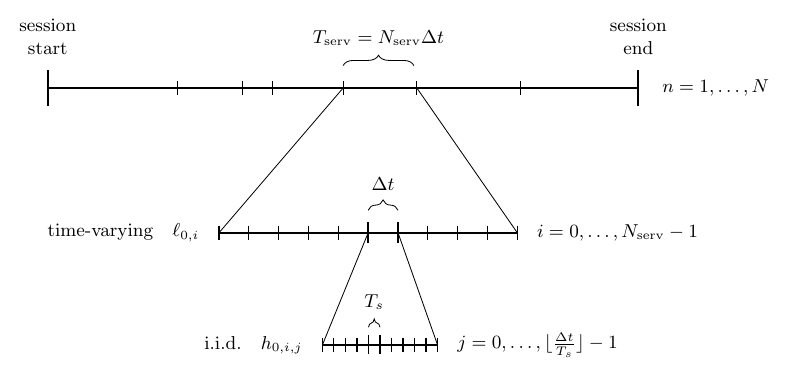}
\caption{{Nested time-scale model within one session interval. A random serving time
$T_{\rm serv}$ between handover events consists of $N_{\rm serv}$ frames of duration
$\Delta t$, where $\Delta t$ denotes the geometric stationarity interval. Each frame is
resolved into $N_s$ symbol intervals of duration $T_s=1/B$ for bandwidth $B$.}}
\label{fig:nestedtimescales}
\end{figure}

The communication link between the user and satellite is modelled as a discrete-time fading AWGN channel with large-scale path-loss and small-scale shadowed-Rician fading. The satellite positions are assumed to be constant within a {geometric stationarity} coherence time { of }$\Delta t$ seconds for the path-loss, and the shadowed-Rician coefficients are assumed to be constant within a coherence time of $T_s$ seconds. In practice, we have that $T_s \ll \Delta t$ since the satellite movement is relatively slow compared to the fast-fading. During the serving time $T_{\rm{serv}} = \mathsf{T}(\Theta_0,\Phi_0,A_0)$, we transmit frames of duration $\Delta t$ seconds equal to the coherence time of the path loss. The resulting number of frames is $N_{\rm{serv}} = \mathsf{N}(\Theta_0,\Phi_0,A_0) = \lfloor 
\mathsf{T}(\Theta_0,\Phi_0,A_0)/\Delta t\rfloor$. 
The maximum number of frames is $N_{\rm{max}} = \lfloor 
T_{\max}/\Delta t\rfloor$ and the minimum number of frames is $N_{\rm{min}} = \lfloor 
T_{\min}/\Delta t\rfloor$. The notation is summarised in Fig. \ref{fig:nestedtimescales}.

The discretised channel output for the $j$-th symbol in the $i$-th frame is
\begin{align}
    y_{i,j} = \frac{{h_{0,i,j}}}{\sqrt{\ell_{0,i}}} x_{i,j} + n_{i,j}
\end{align} 
 for all $j=0,1,\ldots, \lfloor 
\Delta t/T_{s}\rfloor-1$ and $i=0,1,\ldots,N_{\rm{serv}}-1$, where:
\begin{itemize}
    \item $x_{i,j}$ is the complex-valued channel input at time $t=i\Delta t + jT_s$, satisfying the average-power constraint
    $\mathbb{E}\!\left[|x_{i,j}|^{2}\right]\le \gamma$;\footnote{%
    The parameter $\gamma$ is treated as a normalised transmit SNR, independent of specific path-loss contributors (e.g., bandwidth, wavelength, antenna gains, and range), which are omitted from the theoretical analysis.}
    \item $\ell_{0,i} = d^{2}\!\big({\Theta_{0}(i\Delta t),\,\Phi_{0}(i\Delta t)}\big)$ is the free-space LOS path loss;
    \item $h_{0,i,j}$ is an i.i.d.\ complex-valued shadowed-Rician fading process with ergodic parameters $S\!\big(\Theta_{0}({t_{i,j}}),\Phi_0(t_{i,j}), t_{i,j}\big)$ where ${t_{i,j}=i\Delta t + jT_s}$;
    \item $n_{i,j}$ is an i.i.d.\ complex-valued AWGN process with unit variance.
\end{itemize}

{\em Notational convenience:} Within a serving interval, we {sometimes} suppress the intra-frame index $j$ by writing $h_{0,i,j}\equiv h_{0,i}$ and $n_{i,j}\equiv n_i$, {since they are i.i.d. across $j$.} 

\vspace{2mm}
After each serving interval, the handover rule selects a new serving satellite and the channel evolution repeats, yielding an i.i.d.\ renewal structure.

This model admits several natural extensions, e.g., incorporating inter-frame memory when shadowing remains correlated over multiple seconds, or accounting for residual user--satellite interference (if not fully mitigated, as assumed here). Since the subsequent handover analysis relies primarily on the renewal property, we omit these extensions and focus only on assumptions that directly affect the handover strategy.

\subsection{Non-Persistent Satellite Channel}
An interesting special case of the previously defined satellite channel, which models persistent communications, is when the maximum serving time $T_{\max}$ is reduced. As $T_{\max} {\rightarrow \Delta t}$, resulting in $T_{\rm{serv}} {\rightarrow \Delta t}$, transmission only occurs over the initial frame at $i=0$, since $N_{\rm{serv}} {= 1}$. The channel output is therefore
\begin{align}
   y_{0{,j}} = \frac{h_{0,0{,j}}}{\sqrt{\ell_{0,0}}} x_{0{,j}} + n_{{j}} .
\end{align}
Notice that this case completely ignores the part of the propagation process that predicts the orbit trajectory and thus it corresponds to non-persistent communications. Moreover, the semi-stochastic model becomes a purely stochastic model with instantaneous handovers, demonstrating that the persistent channel is a generalisation of the non-persistent channel.

\section{Persistent Satellite Channel Capacity}\label{sec:capacity}

\subsection{Ergodic Capacity: The SatCom Interpretation}\label{sec:ergodic_cap}

Since the ergodic capacity of a persistent satellite channel cannot be evaluated using the typically used formula for ergodic capacity, this section is dedicated to setting up the definition and intuition behind the ergodic capacity for satellite mega-constellations with persistence.

{Let $W(t)$ denote the cumulative delivered payload up to time $t$. As the system undergoes successive handovers, each serving period constitutes one renewal: the channel evolution within a serve depends only on the newly selected satellite, and by construction the rewards and durations are i.i.d.\ across serves. Denoting by $\mathsf{C}\Delta t$ the total capacity (payload) accumulated over a serve (defined explicitly in the next subsection) and by $N_{\rm serv}\Delta t$ the corresponding serve duration, the renewal-reward theorem yields the long-term average throughput
\begin{equation}
    C_{\rm ergodic}
    = \lim_{t\rightarrow\infty}\frac{W(t)}{t} =\frac{\mathbb{E}\!\left[\mathsf{C}\right]}
           {\mathbb{E}\!\left[N_{\rm serv}\right]}\quad\text{(bits/sec)}.
\end{equation}}

The definition above relies on the following modelling assumptions, aligned with the conventional SatCom operating regime:
\begin{enumerate}
    \item No power allocation is performed at the transmitter, resulting in a fixed {transmit SNR}  $\gamma$;
    \item The transmitter and receiver know the path loss $\ell_{0,i}$ using ephemeris data;
    \item The receiver knows the fading coefficient $h_{0,i}$ through channel estimation;
    \item Optimal rate adaptation (ORA) is employed to adjust the coding rate to each instantaneous capacity $W_i$; 
    \item $\Delta t$ is a sufficiently long delay\footnote{This is only required when including the practical constraint of delay much less than the serving times, which is assumed by the SatCom interpretation for practical systems. In theory, this is achieved by coding across multiple serves.} to reliably transmit data at rates close to each instantaneous capacity $W_i$.
\end{enumerate}

In the next section, we instantiate $C_{\text{ergodic}}$ for the persistent satellite channel model introduced earlier: the per-frame rewards are governed by the shadowed-Rician parameters, and the expectation over serving periods is taken with respect to the NBPP-driven handover dynamics.

\subsection{Persistent Capacity}

The capacity of the persistent satellite channel, conditioned on any satellite $(\Theta_k,\Phi_k,A_k) \in \mathcal{V}$ to serve the user for $N_{\rm{serv}}$ frames, is a sum-rate of independent AWGN fading channels with path losses $\{\ell_{k,i}\}$ (which depend on the orbit trajectory $\{\Theta_{k,i},\Phi_{k,i}\}$) and shadowed-Rician coefficients $\{h_{k,i}\}$. Therefore, the total capacity over the serving time is {$\Delta t\mathsf{C}$, with}
\begin{equation}
\begin{aligned}
\mathsf{C}(\Theta_k,\Phi_k, A_k) = \sum_{i=0}^{ {N_{\rm serv}} - 1} C_{k,i},
\end{aligned}
\label{eq:orb_cap}
\end{equation}
where, with shadowed-Rician fading, the instantaneous capacities can be efficiently evaluated as \cite{DiRenzo2010}
\begin{equation}
\begin{aligned}
C_{k,i}&=\mathbb{E}_{|h_{k,i}|^2}\left[\log \left(1 + \frac{\gamma |h_{k,i}|^2}{\ell_{k,i}}\right) \right]\\
&= \frac{1}{\ln 2} \int_{0}^{\infty}
E_i\left(-\frac{s\ell_{k,i}}{\gamma}\right) M^{(1)}(s) ds,
\end{aligned}
\label{eq:rate_integral}
\end{equation}
where $E_i(\cdot)$ is the exponential integral function and $M^{(1)}(s)$ is the first derivative of the shadowed-Rician MGF given by
\begin{equation}
\begin{aligned}
M^{(1)}(s)
&= b_0 (b_0 m)^m (1 + 2b_0 s)^{m - 2} \\
&\quad \times
\frac{
    4 b_0^2 m s + m \Omega + 2b_0 (m + s \Omega)
}{
    \left[b_0\left(m + 2 b_0 m s + s \Omega\right)\right]^{m+1}
} .
\end{aligned}
\label{eq:mgf_derivative}
\end{equation}
Note that this capacity is independent of the handover strategy since we conditioned on an arbitrary serving satellite. 

Let us now average the total capacity with respect to the distribution of the serving satellite, which does depend on the handover strategy, and divide by the average serving time to get the capacity of the persistent satellite channel.

\begin{theorem}[Persistent capacity]\label{thm:pers_cap}
    The ergodic capacity of the persistent satellite mega-constellation channel with handover strategy $\mathsf{H}$ is 
    \begin{align}
        C_{\rm{pers}}[\mathsf{H}] =  \frac{\mathbb{E}_{\mathcal{V},S}\left[ \mathsf{C}(\mathsf{H}(\mathcal{V},S)) \right] }{\mathbb{E}_{\mathcal{V},S}[\mathsf{N}(\mathsf{H}(\mathcal{V},S))]}.
    \end{align}
    where $\mathcal{V}$ is the set of visible satellites from the mega-constellation NBPP with $|\mathcal{V}| \geq 1$.
\end{theorem}
\begin{proof}
     This is a consequence of the renewal reward theorem for renewal reward processes as described in Section \ref{sec:ergodic_cap}.
\end{proof}

\begin{remark}
    The proof of Theorem \ref{thm:pers_cap} is simple to show for the special case where we let the session time $t$ coincide with the duration of exactly $N$ handovers. If we let $\{\mathcal{V}^{(n)}\}$ be the random sets of visible satellites at the handovers $n=1,2,\ldots,N$, which are i.i.d. realisations of the mega-constellation NBPP, then the achievable data rate is
    \begin{align}\label{eq:finite_cap}
        \frac{{\frac{1}{N}}\sum_{n=1}^{N} \mathsf{C}(\mathsf{H}(\mathcal{V}^{(n)},S))}{{\frac{1}{N}}\sum_{n=1}^{N} \mathsf{N}(\mathsf{H}(\mathcal{V}^{(n)},S))} \rightarrow C_{\rm{pers}}[\mathsf{H}]
    \end{align}
    as $N \rightarrow \infty$ due to the strong law of large numbers. When $t$ is arbitrary, we must additionally account for the capacity of the possibly incomplete final serving period. However, since this capacity is bounded, it does not change the limit and the result above remains.
\end{remark}

We remark that while persistent capacity obeys the general form of ergodic capacity \cite{Biglieri1998}, which allows for time-dependence so long as the channel state process is ergodic, it is different to the standard ergodic capacity formula used in the wireless communications literature.

\subsection{Non-Persistent Capacity}\label{sec:C_non_pers}
The capacity of the persistent satellite channel (persistent capacity) as $T_{\max} \rightarrow \Delta t$ yields the capacity of the non-persistent satellite channel (non-persistent capacity). Hence, we have the following corollary of Theorem \ref{thm:pers_cap}, which is the capacity used in the stochastic analysis of handover strategies in \cite{Okati2022b} with the condition $|\mathcal{V}| \geq 1$.

\begin{corollary}[Non-persistent capacity]
    The non-persistent capacity of a satellite mega-constellation with handover strategy $\mathsf{H}$ is
    \begin{align}
        C_{\rm{non-pers}}[\mathsf{H}] =  \mathbb{E}\left[\log \left(1 + \frac{\gamma |h_{0,0}|^2 }{d^2(\Theta_{0},\Phi_{0})}\right) \right] 
    \end{align}
    where $(\Theta_{0},\Phi_{0}, \cdot) = \mathsf{H}(\mathcal{V},S)$.
\end{corollary}
 
The non-persistent capacity is an interesting special case of the persistent capacity since it is an approximate upper bound. {Intuitively, this occurs because shorter serving times give greater flexibility in selecting the most favourable satellite. Formally,} this can be explained using a property of the mutual information $I(\mathbf{x};\mathbf{y})$ between the channel inputs {$\mathbf{x}=[x_{0,1},x_{0,2},\ldots,x_{0,N_{\rm serv}-1}]^{\top}$} and outputs {$\mathbf{y}=[y_{0,1},y_{0,2},\ldots,y_{0,N_{\rm serv}-1}]^{\top}$}, {ignoring CSI for simplicity}, for channels with stationary state processes that satisfy $I(x_{0,0}; y_{0,0})=I(x_{0,i}; y_{0,i})$ for all $i$. For such channels, we have the inequality $I(\mathbf{x}; \mathbf{y}) \leq N_{\rm{serv}} I(x_{0,0}; y_{0,0})$ \cite{Cover2006}, where maximising both sides with respect to the channel inputs gives their respective capacities. In the case of the persistent satellite channel, we have the approximation $I(x_{0,0}; y_{0,0})\approx I(x_{0,i}; y_{0,i})$ whose accuracy depends on how accurately the NBPP models the satellite positions in the deterministic mega-constellation orbits at handover events. As it turns out, it is a very good approximation and serves as an accurate upper bound on the persistent capacity for a given handover strategy. 

\subsection{Upper Bound on Persistent Capacity}
The persistent capacity $C_{\rm pers}[\mathsf{H}]$ depends on the handover strategy $\mathsf{H}$ that is chosen. In the next section, we will derive the handover strategy that maximises this capacity. However, since numerically computing $C_{\rm pers}[\mathsf{H}]$ is difficult in general, due to the order statistics often required to describe a handover strategy, it is useful to at least have capacity bounds that can be numerically computed. The following proposition provides a closed-form upper bound that is independent of the handover strategy. We note that this upper bound cannot be achieved by any realisable handover strategy. In the next section, we will pair this upper bound with a lower bound based on the worst-case handover strategy.

\begin{proposition}\label{prop:C_UB}
    For an arbitrary handover strategy $\mathsf{H}$, which need not obey Definition \ref{def:HO}, we have the upper bound
    \begin{align}
        \overline{C}_{\rm pers} = \max_{a\in\{-1,+1\}}\sup_{(\theta,\phi)\in\mathsf{Cap}} \left\{\frac{\mathsf{C}(\theta,\phi,a)}{\mathsf{N}(\theta,\phi,a)}\right\}
    \end{align}
    that satisfies $C_{\rm pers}[\mathsf{H}]\leq \overline{C}_{\rm pers}$.
\end{proposition}
{
\begin{proof}
For all $(\theta,\phi,a)$ with $\mathsf{N}(\theta,\phi,a)>0$, $\mathsf{C}(\theta,\phi,a)/\mathsf{N}(\theta,\phi,a) \le\overline{C}_{\rm pers}$ gives $
\mathsf{C}(\theta,\phi,a) \le \overline{C}_{\rm pers}\,\mathsf{N}(\theta,\phi,a).$
Taking expectations over $(\theta,\phi,a)$ yields $\mathbb{E}[\mathsf{C}] \le \overline{C}_{\rm pers}\,\mathbb{E}[\mathsf{N}]$, and dividing by $\mathbb{E}[\mathsf{N}]>0$ gives the claim.
\end{proof}}

\section{Optimal Handover Strategy}\label{sec:handover}

For a fixed handover strategy $\mathsf{H}$, the persistent capacity was derived in Theorem \ref{thm:pers_cap} as the maximum achievable rate. We now consider the maximisation of persistent capacity with respect to the handover strategy to find the optimal handover strategy.

\begin{definition}[Optimal handover strategy]
    The handover strategy $\mathsf{H}^{\star}$ is optimal if
    \begin{align}
     C_{\rm{pers}}[\mathsf{H}^{\star}] = \sup_{\mathsf{H}\in \mathcal{H}} C_{\rm{pers}}[\mathsf{H}]
    \end{align}
    where $\mathcal{H}$ is the set of all handover strategies that satisfy Definition \ref{def:HO}.
\end{definition}

\subsection{Dinkelbach's Algorithm}
For convenience, denote $C_{k,n}$ as the total capacity, $N_{k,n}$ as the number of frames in the serving period, and $p_{k,n}$ as a binary variable indicating the handover decision, at the $n$-th handover event of the $k$-th visible satellite in $\mathcal{V}^{(n)}$. {Define the vectors $\mathbf{C}_n = [C_{1,n}, \ldots, C_{|\mathcal{V}^{(n)}|,n}]^\top$, $\mathbf{N}_n = [N_{1,n}, \ldots, N_{|\mathcal{V}^{(n)}|,n}]^\top$, and $\mathbf{p}_n = [p_{1,n}, \ldots, p_{|\mathcal{V}^{(n)}|,n}]^\top$. Then, the normalised capacity can be expressed compactly as
\begin{align}
    C_N(\mathbf{p})
    = \frac{\sum_{n=1}^{N} \mathbf{p}_n^\top \mathbf{C}_n}
           {\sum_{n=1}^{N} \mathbf{p}_n^\top \mathbf{N}_n}
    = \frac{U_N(\mathbf{p})}{V_N(\mathbf{p})},
\end{align}}
and then solving for the optimal handover decisions $\mathbf{p}$ is the $0$-$1$ fractional program~\cite{Matsui1992} 
\begin{align}
C^* &= \max_{\mathbf{p}\in  \mathcal{P}} C_N(\mathbf{p}),
\end{align}
where $\mathcal{P} = \{\mathbf{p} = (\mathbf{p}_1, \ldots, \mathbf{p}_N) : \mathbf{p}_n \in \{0,1\}^{|\mathcal{V}^{(n)}|}, \; \mathbf{1}^\top \mathbf{p}_n = 1 \text{ for all } n\}$. While this is a non-convex program in general, Dinkelbach's algorithm (\cite{Dinkelbach1967}, Algorithm \ref{alg:Dinkelbach}) transforms it into a sequence of convex programs whose solutions converge to the global optimum. In particular, the Dinkelbach transform replaces the maximisation of the non-convex objective $C_N(\mathbf{p})$ with the maximisation 
\begin{align}\label{eq:Dinkelbach_transform}
    F_N(C) = \frac{1}{N}\max_{\mathbf{p}\in \mathcal{P}} \{U_N(\mathbf{p})-C V_N(\mathbf{p}) \},
\end{align}
where $C$ is an initial guess of the maximum capacity $C^*$ at the global optimum $\mathbf{p}^*$. After solving $F_N(C)$ with some guess $C$ to get a solution $\mathbf{p}$, the guess is updated as  $C_N(\mathbf{p})$. Dinkelbach showed that $F_N(C)$ monotonically decreases until $U_N(\mathbf{p})-C V_N(\mathbf{p}) = 0$ is satisfied{, then} $C=C^*$ and $\mathbf{p}=\mathbf{p}^*$.

Observe that the handover decisions $\mathbf{p}$ are general in that they do not necessarily correspond to a valid handover strategy as in Definition \ref{def:HO}, which only allows memoryless handover decisions. This could result in an undefined capacity since the processes $\{U_N(\mathbf{p}^*)\}$ and $\{V_N(\mathbf{p}^*)\}$ need not be ergodic and consequently $C_N(\mathbf{p}^*)$ may not converge to the persistent capacity in Theorem \ref{thm:pers_cap} as $N \rightarrow\infty$. This issue will be addressed by modifying Dinkelbach's transform.

{
The time-complexity of the standard Dinkelbach algorithm, shown in Algorithm~\ref{alg:Dinkelbach}, 
has been extensively analysed. In~\cite{Matsui1992}, the number of iterations required for 0--1 
fractional programs is \(O(\log N)\) iterations, corresponding to {\em superlinear} convergence. 
Consequently, the dominant computational cost arises from the integer linear-program solver executed 
at each iteration, scaled by the number of iterations to give the overall time-complexity.
}

\begin{algorithm}
\caption{Dinkelbach's algorithm}\label{alg:Dinkelbach}
\begin{algorithmic}[1]
\State Set $C=C_0$ as an initial guess $C_0$.
\State Solve the linear program $F_N(C)$ to obtain an optimal solution $\mathbf{p}$.
\State Update guess $C$ with $C_N(\mathbf{p})$
\State If $F_N(C) = 0$ then return $C$ else go to Line 2.
\end{algorithmic}
\end{algorithm}

\subsection{Dinkelbach-type Algorithm for Optimal Handover}
In this section, we consider a simplification of Dinkelbach's algorithm for numerator and denominator coefficients of the objective that are i.i.d. processes. In particular, if we can restrict $\mathbf{p}$ to follow Definition \ref{def:HO}, then $\{U_N(\mathbf{p})\}$ and $\{V_N(\mathbf{p})\}$ are i.i.d. processes such that 
\begin{align}\label{eq:cap_lim}
    C_{\rm{pers}}[\mathsf{H}]&=\lim_{N\rightarrow\infty} C_N(\mathbf{p}),
\end{align}
by the strong law of large numbers, where $\mathbf{p}$ now corresponds to some valid $\mathsf{H} \in \mathcal{H}$. Under this uncoordinated handover regime, Dinkelbach's solution is asymptotically equivalent to the following optimal handover strategy.

\begin{theorem}\label{thm:opt_HO}
    The optimal handover strategy chooses a visible satellite from $\mathcal{V}$ with fading parameters determined by $S$ as
\begin{equation}
\begin{aligned}
\mathsf{H}^{\star}(\mathcal{V}, S) 
&= \argmax_{(\Theta_k, \Phi_k, A_k) \in \mathcal{V}} \Big\{ 
\mathsf{C}(\Theta_k, \Phi_k, A_k) \\
&\qquad\qquad\qquad\quad - C_{\rm{pers}}[\mathsf{H}^{\star}] \, \mathsf{N}(\Theta_k, \Phi_k, A_k) \Big\} .
\end{aligned}
\label{eq:opt_HO}
\end{equation}
\end{theorem}

\begin{proof}
Let $Q^{(n)}(C) = \max_{1 \leq k \leq |\mathcal{V}^{(n)}|}
\{ C_{k,n} - C N_{k,n} \}$ and
$Q_N(C) = \frac{1}{N}\sum_{n=1}^{N} Q^{(n)}(C)$. Observe that
\begin{equation}
\begin{aligned}
Q(C)&= \lim_{N \rightarrow \infty} Q_N(C) \\
&=\mathbb{E}_{\mathcal{V}, S} \left[\max_{1\leq k \leq |\mathcal{V}|}
\{ \mathsf{C}(\Theta_k,\Phi_k,A_k)
- C \mathsf{N}(\Theta_k,\Phi_k,A_k)\}\right]
\end{aligned}
\label{eq:Q_def}
\end{equation}
is the limiting Dinkelbach residual for memoryless, uncoordinated handovers.
{The residual \(Q_N(C)\) is strictly decreasing and has a unique
zero by Lemmas~\ref{lem:QN_monotone} and~\ref{lem:QN_zero}; the same argument
applies to \(Q(C)\). Let \(C^\star\) denote the unique zero of \(Q(C)\)}.

{We now show that this zero is the optimal ratio of expectations. For any
admissible memoryless handover strategy \(\mathsf H\), its selected satellite
cannot have a larger residual than the maximising satellite in \eqref{eq:Q_def}.
Hence,
\begin{align}
&\mathbb{E}_{\mathcal V,S}\!\left[
\mathsf C(\mathsf H(\mathcal V,S))
-
C^\star\mathsf N(\mathsf H(\mathcal V,S))
\right] \leq Q(C^\star)=0 .\notag
\end{align}
Since \(\mathbb{E}_{\mathcal V,S}[\mathsf N(\mathsf H(\mathcal V,S))]>0\),
this implies \(C_{\rm pers}[\mathsf H]\leq C^\star\). Equality is achieved by
$\mathsf{H}^\star$ in \eqref{eq:Q_def} with \(C=C^\star\). Therefore
\(C^\star=C_{\rm pers}[\mathsf H^\star]\), and the globally optimal
memoryless handover rule is given by \eqref{eq:opt_HO}.}
\end{proof}

That is, assuming that the maximum capacity is known, the optimal satellite for handover is selected by maximising the residual between the instantaneous {serving} capacity and the averaged maximum capacity. To compute the maximum capacity \(C_{\mathrm{pers}}[\mathsf{H}^{\star}]\), Dinkelbach’s algorithm is modified to obtain the simplified Dinkelbach-type procedure described in Algorithm~\ref{alg:max_pers_cap}, where the key distinction is that the maximisation is performed over pairs of numerator and denominator terms, rather than jointly over all terms.

The advantage of Algorithm~\ref{alg:max_pers_cap} over Algorithm~\ref{alg:Dinkelbach} 
is that its maximisation step {separates across handover events, so each 
handover decision only requires maximisation over the visible satellites.} 
However, the number of iterations required for convergence in 
Algorithm~\ref{alg:max_pers_cap} is nontrivial to characterise, {since it is a variant of Dinkelbach's algorithm.} 
In addition, the {decision complexity} of \(\mathsf H^{\star}\) is \(O(N_{\rm vis}N_{\max})\) after \(C_{\rm pers}[\mathsf H^{\star}]\) has been 
estimated, as summarised in Table~\ref{tab:model_scope_complexity}.

Interestingly, the optimal handover strategy can be applied to more general channel models with arbitrary  serving capacity functions $\mathsf{C}$, assuming the capacity limit in (\ref{eq:cap_lim}) exists. {For example, the algorithm can be applied to an operational mega-constellation to iteratively estimate $C_{\rm pers}[\mathsf{H}^\star]$ from the most recent $N$ handovers. If the estimate is formed using a sliding window or a forgetting factor, then---while the decision rule remains optimal for fixed $S$---the resulting strategy becomes adaptive to slowly time-varying $S$.}

\begin{algorithm}
\caption{Dinkelbach-type algorithm for estimating $C_{\rm pers}[\mathsf{H}^{\star}]$}\label{alg:max_pers_cap}
\begin{algorithmic}[1]
\State Set $C=C_0$ as an initial guess $C_0$.
\State Solve $Q^{(n)}(C)$, $1\leq n\leq N$, to obtain solution $\mathbf{p}$.
\State Update guess $C$ with $C_N(\mathbf{p})$.
\State If $Q_N(C) < \epsilon$ then return $C$ else go to Line 2.
\end{algorithmic}
\end{algorithm}

If we wish to have a handover strategy that does not require knowledge of $C_{\rm{pers}}[\mathsf{H}^{\star}]$, as in the optimal handover strategy, then we could instead maximise the capacity over one serve as in the {\em max. serving capacity (MSC)} handover strategy \cite{Mcbain2025b}
    \begin{align}\label{eq:MOC_HO}
         \mathsf{H}_{\rm MSC}(\mathcal{V},S)=\argmax_{(\Theta_k,\Phi_k,A_k)\in\mathcal{V}} \left\{ \frac{\mathsf{C}(\Theta_k,\Phi_k,A_k)}{\mathsf{N}(\Theta_k,\Phi_k,A_k)} \right\}.
    \end{align}
 For the case of constant serving times, which arises when $T_{\min}=T_{\max}$, then we have $\mathsf{H}_{\rm MSC} = \mathsf{H}^{\star}$. 
{
Moreover, they may be approximately equal when capacity is insensitive to serving time due to slowly varying satellite path losses.
} Therefore, this series of simplifications shows how the general optimal handover strategy based on the persistent capacity relates to the existing handover strategies in the literature. This highlights the cases in which each of the previously proposed strategies are optimal, depending on the assumptions or requirements of the handover.

\subsection{Non-Persistent Handover}
Now let us consider the non-persistent scenario where $C_{\rm{pers}}[\mathsf{H}]=C_{\rm{non-pers}}[\mathsf{H}]$. This scenario has a constant serving time of $1$ sample, and therefore the optimal handover strategy is to maximise the orbit capacity as in the handover strategy
\begin{equation}
\begin{aligned}
&\mathsf{H}_{\rm MSC_0}(\mathcal{V}, S) \\
&= \argmax_{(\Theta_k, \Phi_k, \cdot) \in \mathcal{V}} 
\left\{ \mathbb{E}_{|h_{k,0}|^2} \left[
\log\left( 1 + \frac{\gamma |h_{k,0}|^2}{d^2(\Theta_k, \Phi_k)} \right) \right] \right\} .
\end{aligned}
\label{eq:MOC_0_HO}
\end{equation}
Since there is no orbit trajectory prediction in this scenario, this strategy may be improved by removing the average with respect to the fading coefficients $\{h_{k,0}\}$ if they are assumed to be side-information, since they can theoretically be estimated close to the handover time; this is the handover strategy that was employed in \cite{Okati2022b}.
If the fading coefficients are identical as $h_{1,0}=h_{2,0}=\cdots=h_{N_{\rm vis},0}$, which coincides with the case of no fading if they equal a constant of $1$, then we can equivalently minimise distance as in the nearest-satellite handover strategy; in this scenario, the capacity upper bound in Proposition \ref{prop:C_UB} is the Shannon capacity $\log(1+\gamma/h^2)$ with a best-case path loss $h^2$.

\subsection{Worst-Case Handover without Side-Information}
As a benchmark for the other more sophisticated handover strategies, we now consider a worst-case handover strategy. Let us restrict the CCU to not be allowed to use any side-information about the visible satellites $\mathcal{V}$ and fading parameters $S$ {for making} the handover decision. In this scenario, the ordering of the visible satellites in $\mathcal{V}$ is arbitrary and hence the best handover strategy is to choose a random satellite as
\begin{align}
    \mathsf{H}_{\rm Rand} = \mathcal{U}(\mathcal{V}) .
\end{align}

This handover strategy is particularly tractable for analysis since it does not involve order statistics. In fact, the persistent capacity with this handover strategy can be evaluated via numerical integration using \cite[Lemma 1]{Mcbain2025b}; for completeness, we summarise this Lemma and its proof in Appendix \ref{appendix:C_rand_HO}. Note that the previous handover strategies include order statistics that make numerical computation of persistent capacity very challenging, unless the serving times are sufficiently short such that it equals the non-persistent capacity from Section \ref{sec:C_non_pers}.

\begin{proposition}
Handover strategies satisfying Definition~\ref{def:HO} that do not use
satellite-specific information from $\mathcal{V}$ or $S$ have the same
persistent capacity as the random handover strategy, giving $C_{\rm pers}[\mathsf{H}]
=
C_{\rm pers}[\mathsf{H}_{\rm Rand}]$.
\end{proposition}

{
\begin{proof}
Conditioned on $N_{\rm vis}=n$, the satellite states in $\mathcal{V}$ are
exchangeable. If $\mathsf{H}$ does not use the satellite states in
$\mathcal{V}$ or the fading parameters $S$, its selected index is
independent of the state associated with that index. Consequently, the
selected satellite under $\mathsf{H}$ has the same distribution as a
uniformly selected satellite under $\mathsf{H}_{\rm Rand}$. Hence, $C_{\rm pers}[\mathsf{H}]
=
C_{\rm pers}[\mathsf{H}_{\rm Rand}]$.
\end{proof}}

{To summarise the modelling assumptions and practical requirements of the
handover strategies, Table~\ref{tab:model_scope_complexity} compares the
required information, capacity evaluation, and decision complexity of each
strategy. }

\begin{table*}[t]
\centering
\caption{{Model scope, information requirements, and decision complexity of the considered handover strategies.}}
\label{tab:model_scope_complexity}
\scriptsize
\setlength{\tabcolsep}{3pt}
\renewcommand{\arraystretch}{1.12}
\begin{tabular}{
p{0.10\textwidth}
p{0.25\textwidth}
p{0.25\textwidth}
p{0.18\textwidth}
p{0.13\textwidth}
}
\toprule
Strategy
& Required information
& Strategy-specific assumption
& Decision metric
& Decision complexity \\
\midrule

$\mathsf H_{\rm Rand}$
& Visible set $\mathcal V$.
& No candidate or fading side-information.
& Random candidate.
& $O(1)$ \\

$\mathsf H_{\rm MSC_0}$
& Current positions and instantaneous or statistical CSI.
& No orbit prediction.
& Initial frame capacity.
& $O(N_{\rm vis})$ \\

$\mathsf H_{\rm MSC}$
& Candidate positions, serving times, and fading statistics.
& Orbit prediction available.
 & \begin{tabular}[t]{@{}l@{}}
  Serving capacity\\
  $\mathsf C/\mathsf N$.
  \end{tabular}
& $O(N_{\rm vis}N_{\rm max})$ \\

$\mathsf H^{\star}$
& Candidate positions, serving times, fading statistics, and
$C_{\rm pers}[\mathsf H^\star]$.
& Optimal for memoryless handover.
& \begin{tabular}[t]{@{}l@{}}
  Dinkelbach residual\\
    $\mathsf C-C_{\rm pers}[\mathsf H^\star]\mathsf N$.
  \end{tabular}
& $O(N_{\rm vis}N_{\rm max})$\\
\midrule
\addlinespace[1mm]
\multicolumn{5}{p{0.96\textwidth}}{\scriptsize
All strategies are user-centric, memoryless, and uncoordinated; residual interference is
treated as noise; and serving periods are independent under the semi-stochastic
renewal model. Load-coupled multi-user association, coordinated interference
management, and prediction errors in orbit or channel information are outside
the optimisation problem.}\\
\bottomrule
\end{tabular}
\end{table*}

\section{Numerical Results}\label{sec:numerical_results}

The numerical results to come in this section will compare the handover strategies $\mathsf{H}_{\rm Rand}$, $\mathsf{H}_{\rm MSC_0}$, $\mathsf{H}_{\rm MSC}$, and $\mathsf{H}^{\star}_{\rm }$, as defined in the previous section, in terms of the persistent capacity from Theorem \ref{thm:pers_cap}. The persistent capacity with $\mathsf{H}_{\rm Rand}$ is computed using numerical integration to give a lower bound, paired with the upper bound from Proposition \ref{prop:C_UB}. For the other handover strategies, we use Monte Carlo simulations using the sampling procedure in Theorem \ref{thm:sampling_thm} with $10^3$ realisations of the mega-constellation to accurately estimate the persistent capacities. To verify the semi-stochastic model for evaluating handover strategies, we use the Starlink mega-constellation from a TLE file with an epoch date of 2024/01/01, filtering the satellites with parameters that satisfy $b=53 \pm 1^\circ$ and $h = 550 \pm 50$ km. The mega-constellation is simulated to get satellite positions at $\Delta t = 1$ second intervals, for a duration over $80$ hours with SGP4 orbit propagation and for $10$ orbital periods with circular orbit (CIRC) propagation. The position data from these simulations is used to estimate the persistent capacities according to the SGP4 and CIRC models for comparison with the theoretical NBPP model. The shadowed-Rician fading model is parameterised using the ``average shadowing'' parameters from \cite[Table III]{Abdi2003}, which correspond to choosing $S(\theta,\phi; t)=(b_0,m,\Omega) = (0.126, 10.1,0.835)$.\footnote{Alternatively, we could upgrade the model to include time-dependent satellite positions using \cite[Eq. 19]{Abdi2003} to get $S(\theta(t),\phi(t), t)=(b_0(\sigma(t)),m(\sigma(t)),\Omega(\sigma(t)))$. {We omit this here to keep our results comparable with results in the literature}.}
Since the handover strategies are user-centric, we must also choose a ground user location: we study ground users in Melbourne, {Australia}, and in Helsinki, {Finland}, which represent {distinct} locations with a low and a high latitude, respectively.

\subsection{Serving Capacity}
Recall that the sub-optimal max. serving capacity handover strategy $\mathsf{H}_{\rm MSC}$ maximised the serving capacity $\mathsf{C}(\theta,\phi,a)/\mathsf{N}(\theta,\phi,a)$ over all visible satellite parameters in $\mathcal{V}$, and additionally recall that the persistent capacity could be upper bounded by maximising the serving capacity over $\mathsf{Cap}$ (i.e., without the restriction of a finite set of visible satellites). Since the metric for $\mathsf{H}_{\rm MSC}$ directly relates to $C_{\rm pers}$, it is a particularly interesting function to study that provides insights on the qualitative features considered when choosing a serving satellite.

\begin{figure*}
    \centering
    \begin{subfigure}[b]{0.45\textwidth}
        \centering
        \includegraphics[width=1\textwidth]{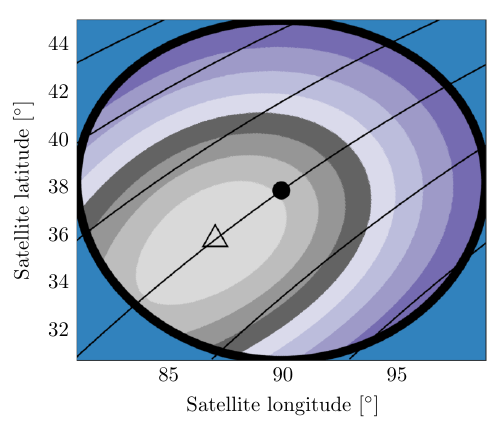}
        \caption{\centering Ground user in Melbourne \protect\linebreak $(\psi_{\min}=30^{\circ},T_{\rm min}=0, T_{\max}=\infty)$}
    \end{subfigure}%
    ~ 
    \begin{subfigure}[b]{0.45\textwidth}
        \centering
        \includegraphics[width=1\textwidth]{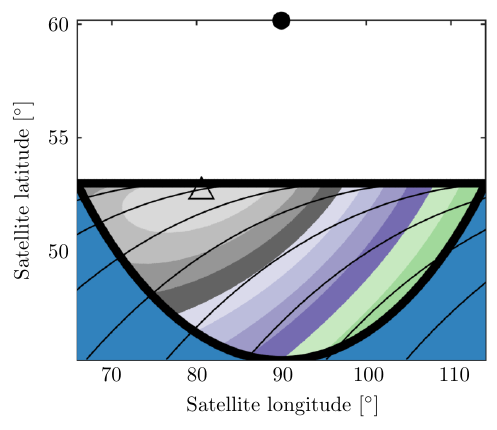}
        \caption{\centering Ground user in Helsinki \protect\linebreak $(\psi_{\min}=10^{\circ},T_{\rm min}=0, T_{\max}=\infty)$}
    \end{subfigure}
        ~ 
    \begin{subfigure}[b]{0.45\textwidth}
        \centering
        \includegraphics[width=1\textwidth]{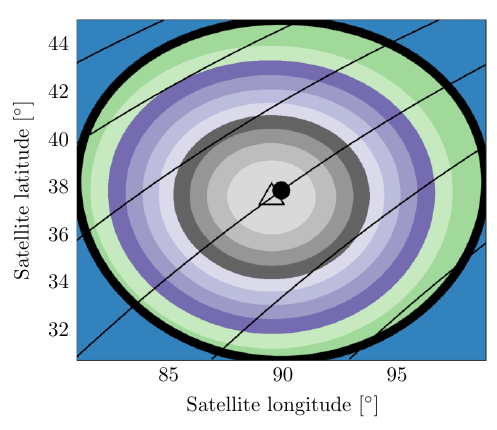}
        \caption{\centering Ground user in Melbourne \protect\linebreak $(\psi_{\min}=30^{\circ},T_{\rm min}=T_{\max}=15\text{s})$}
    \end{subfigure}%
        ~ 
    \begin{subfigure}[b]{0.45\textwidth}
        \centering
        \includegraphics[width=1\textwidth]{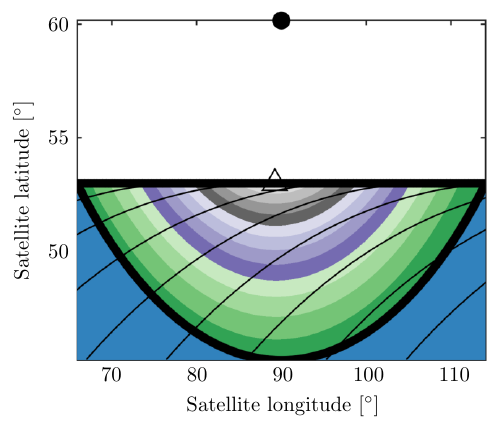}
        \caption{\centering Ground user in Helsinki \protect\linebreak $(\psi_{\min}=10^{\circ},T_{\rm min}=T_{\max}=15\text{s})$}
    \end{subfigure}
    \caption{Heat maps of the serving capacity $\mathsf{C}(\theta,\phi,1)/\mathsf{N}(\theta,\phi,1)$ for ascending satellites. Light grey is the highest serving capacity, dark green is the lowest serving capacity, and blue (and white) is zero capacity (outside the visibility cap). The black dot is the location of the ground user, the triangle marker is the satellite location that achieves the capacity upper bound $\overline{C}_{\rm pers}$, the thin black lines are example orbit trajectories, and the thick black curve is the boundary of the visibility cap $\mathsf{Cap}$.}
    \label{fig:ServCap}
\end{figure*}

In Fig. \ref{fig:ServCap}, the serving capacity for ascending satellites is plotted as a heat map for each ground user without serving time constraints and with a fixed serving time of $15$ seconds. For each case, we plot the satellite position with the highest reliability, corresponding to $\overline{C}_{\rm pers}$. Without serving time constraints in Fig. \ref{fig:ServCap}(a) and Fig. \ref{fig:ServCap}(b), we observe that the most reliable serving satellites are those that have recently entered the visibility cap and will eventually pass nearby the user; the least reliable satellites are those that are leaving the visibility cap, since they are moving further away from the user. For the fixed serving time in Fig. \ref{fig:ServCap}(c) and Fig. \ref{fig:ServCap}(d), we observe that the serving capacity follows $\sigma(\theta,\phi)\propto d(\theta,\phi)$, with a slight skew towards the bottom-left due to the persistence introduced by the $15$-second serving time. This may be viewed as a visual proof that {such short durations} are almost without persistence, and thus we can essentially just choose the nearest satellite. Note that the shadowing model is not causing additional skewing in this example, since the shadowing parameters chosen for $S$ are independent of $(\theta,\phi)$.

\subsection{Persistent Capacity without Serving Time Constraints ($T_{\min}=0$, $T_{\max}=\infty$)}

Let us now consider the case of unconstrained serving times, which corresponds to a minimum serving time $T_{\min} = 0$ and a maximum serving time $T_{\max}=\infty$. In this scenario, the only constraint is that we can only be served for the serving satellite's visibility time, {so that} $T_{\rm serv} = T_{\rm vis}(\Theta_0,\Phi_0,A_0)$, i.e., while the satellite is in the visibility cap. The motivation for studying this scenario is that it results in long serving times {and therefore a low handover rate}, since the satellite persistence is high. Since this results in variable serving times, it is a key example for demonstrating the performance of $\mathsf{H}^{\star}_{\rm }$ relative to the other sub-optimal handover strategies. 

\begin{figure*}
    \centering
    \begin{subfigure}[b]{0.5\textwidth}
        \centering
        \includegraphics[width=1\textwidth]{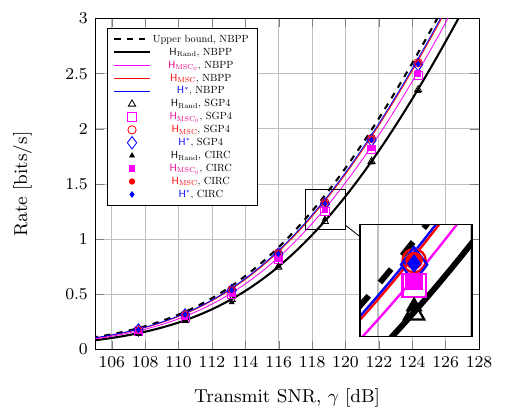}
        \caption{Ground user in Melbourne $(\psi_{\min}=30^{\circ})$}
    \end{subfigure}%
    ~ 
    \begin{subfigure}[b]{0.5\textwidth}
        \centering
        \includegraphics[width=1\textwidth]{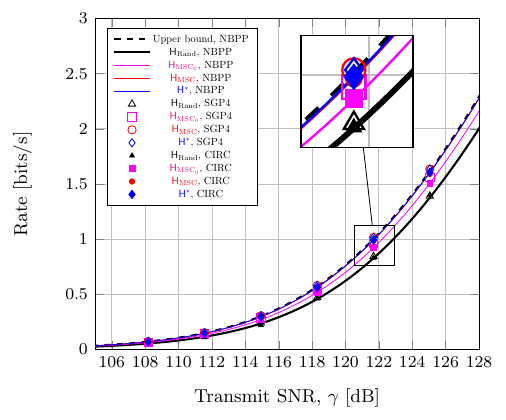}
        \caption{Ground user in Helsinki $(\psi_{\min}=10^{\circ})$}
    \end{subfigure}
    \caption{Persistent capacity with cap serving times for a range of {transmit SNR}s.}
    \label{fig:C_cap_serv}
\end{figure*}

\begin{figure*}
    \centering
    \begin{subfigure}[b]{0.49\textwidth}
        \centering
        \includegraphics[width=\textwidth]{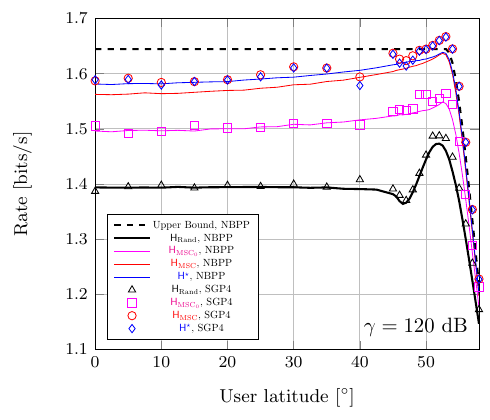}
        \caption{Varying user latitude with $\psi_{\min}=30^{\circ}$.}
        \label{fig:C_cap_serv2_lat}
    \end{subfigure}
    \hfill
    \begin{subfigure}[b]{0.49\textwidth}
        \centering
        \includegraphics[width=\textwidth]{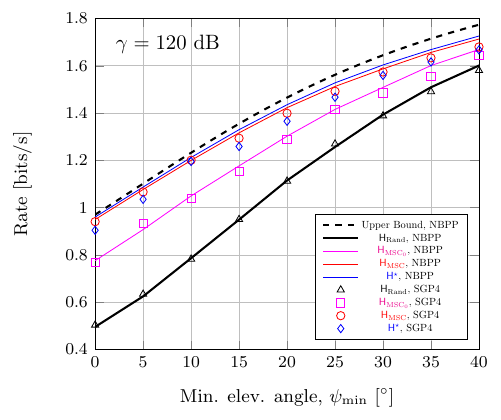}
        \caption{Varying minimum elevation angle for a ground user in Melbourne.}
        \label{fig:C_cap_serv2_psimin}
    \end{subfigure}
    \caption{{Persistent capacity sensitivity to user latitude and minimum elevation angle.}}
    \label{fig:C_cap_serv2}
\end{figure*}

In terms of modelling accuracy, Fig.~\ref{fig:C_cap_serv} {shows} a close agreement between the persistent capacities for NBPP, SGP4 and CIRC. This supports the accuracy of the semi-stochastic model and suggests the handover independence assumption is a {reasonable} approximation. 
{
Notably, $C_{\rm pers}[\mathsf{H}^{\star}]$ exhibits the largest discrepancy between the NBPP model and the SGP4/CIRC simulations, {which is likely due the independence assumption}. For the Melbourne user, $\mathsf{H}^{\star}$ in fact underperforms $\mathsf{H}_{\rm MSC}$. Consequently, the incremental improvement predicted for $\mathsf{H}^{\star}$ over $\mathsf{H}_{\rm MSC}$ is unlikely to be operationally significant. {The candidate satellite that maximises total serving capacity also almost always maximises the instantaneous capacity rate and the Dinkelbach-style persistent-capacity objective. Therefore, in this regime, the capacity versus serving time tradeoff is weak, and serving-time effects do not substantially alter the handover decision.} We therefore view $\mathsf{H}^{\star}$ not as a recommended strategy, but as a theoretical benchmark that clarifies the structure of the problem and motivates the design of $\mathsf{H}_{\rm MSC}$ through its relationship to $\mathsf{H}^{\star}$. {In practice, this is advantageous because $\mathsf{H}_{\rm MSC}$ is substantially simpler while achieving essentially the same persistent capacity as the optimal rule in the real-data constellation.}
}
In addition, it is interesting to note that $C_{\rm pers}[\mathsf{H}_{\rm Rand}]$ has the closest {model agreement} for both users. {We also note that Algorithm~\ref{alg:max_pers_cap} converges within five iterations in these examples.}

Comparing the handover strategies in Fig.~\ref{fig:C_cap_serv}, we confirm that the handover strategies with more side-information have higher persistent capacities for both users. For both users, $\mathsf{H}_{\rm MSC_0}$ has a gain of $\{0.62, 0.67\}$ dB over $\mathsf{H}_{\rm Rand}$, $\mathsf{H}_{\rm MSC}$ has a gain of $\{0.38, 0.45\}$ dB over $\mathsf{H}_{\rm MSC_0}$, and $\mathsf{H}^{\star}_{\rm }$ has a gain of $\{0.07, 0.03\}$ dB over $\mathsf{H}_{\rm MSC}$. Overall, the optimal handover strategy has a gain of $\{1.07, 1.15\}$ dB compared to the worst-case handover strategy. That is, by using all of the available information in the handover strategy, we have effectively increased the transmit power by more than $1$ dB for both users.

The capacity upper bound $\overline{C}_{\rm pers}$ in Proposition~\ref{prop:C_UB} is observed to be tight in Fig.~\ref{fig:C_cap_serv}, suggesting that there is often {a visible satellite} in the light grey regions of {the} respective heat maps in Fig.~\ref{fig:ServCap}(a) and Fig.~\ref{fig:ServCap}(b). In addition, the bound is especially tight for the higher latitude user since the density of satellites according to $f_{\Phi}(\phi)$ is higher. For such serving times, this upper bound may actually be sufficiently tight for use in a simplified theoretical analysis.

\begin{figure*}
    \centering
    \begin{subfigure}[b]{0.5\textwidth}
        \centering
        \includegraphics[width=1\textwidth]{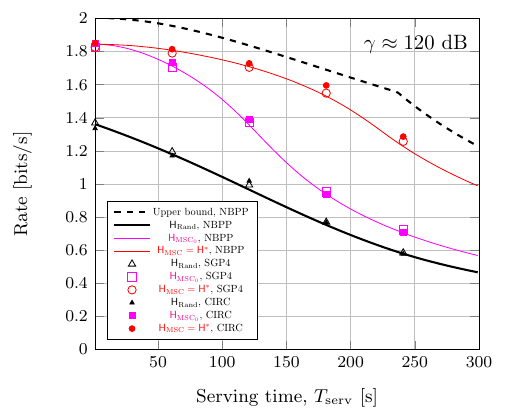}
        \caption{Ground user in Melbourne $(\psi_{\min}=30^{\circ})$}
    \end{subfigure}%
    ~ 
    \begin{subfigure}[b]{0.5\textwidth}
        \centering
        \includegraphics[width=1\textwidth]{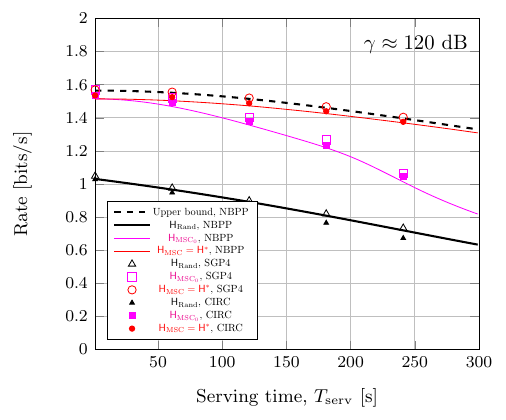}
        \caption{Ground user in Helsinki $(\psi_{\min}=10^{\circ})$}
    \end{subfigure}
    \caption{Persistent capacity for a range of fixed serving times and a fixed {transmit SNR} $\gamma \approx 120$ dB.}
    \label{fig:C_fixed}
\end{figure*}

{
In Fig.~\ref{fig:C_cap_serv2}, we fix the transmit SNR to $120$~dB and examine the sensitivity of persistent capacity to user latitude and minimum elevation angle. For fixed $\psi_{\min}=30^{\circ}$, the persistent capacity roughly follows the visible-satellite density as the user latitude varies: it remains nearly flat at low latitudes, increases toward the high-density latitude region, and then decreases rapidly near the edge of coverage. For a fixed ground user in Melbourne, the persistent capacity increases steadily with the minimum elevation angle, since the visibility cap restricts the candidate satellites to those closer to the maximum instantaneous-capacity region, indicated by the grey regions in Fig.~\ref{fig:ServCap}. However, the model accuracy relative to SGP4 appears to degrade for small visibility caps, where fewer satellites are visible on average. Overall, the relative gaps between handover strategies remain similar to those observed in Fig.~\ref{fig:C_cap_serv}. Moreover, $\mathsf{H}_{\rm MSC}$ remains slightly better overall under SGP4. Interestingly, the regime in which $\mathsf{H}_{\rm MSC}$ under SGP4 outperforms $\mathsf{H}_{\rm MSC}$ under NBPP is also the regime in which it essentially matches $\mathsf{H}^{\star}$.

\subsection{Persistent Capacity with Fixed Serving Times ($T_{\min}=T_{\max}$)}

Let us now consider the case of fixed serving times, which corresponds to constraining the minimum and maximum serving times as $T_{\min}=T_{\max}=T_{\rm serv}$ where $T_{\rm serv}$ is now a constant that we can set. Unlike with the cap serving times, in this scenario $\mathsf{H}^{\star}_{\rm }$ coincides with $\mathsf{H}_{\rm MSC}$ since the serving times are constant. A strong motivation for studying fixed serving times is that the Starlink mega-constellation network is known to perform inter-satellite handovers every 15 seconds \cite{ZhaoPan2024}, which are synchronised to occur at the 12th, 27th, 42nd, and 57th second past every minute for all users.  

As earlier, we observe in Fig. \ref{fig:C_fixed} that the persistent capacities for all handover strategies remain in close agreement between NBPP, SGP4, and CIRC with the additional constraint of fixed serving times. We note that the data points for SGP4 and CIRC with longer serving times are less accurate, since there are less handovers over the same simulation period to average over, but this appears to be insignificant in the results. 

As expected, we observe in Fig. \ref{fig:C_fixed} that $\mathsf{H}_{\rm MSC_0}$ and $\mathsf{H}_{\rm MSC}$ have an equal persistent capacity for short serving times, which is $1.8442$ bits/s in Melbourne and reduces to $1.5145$ bits/s in Helsinki due to the higher latitude. In addition, $\mathsf{H}_{\rm Rand}$ degrades the capacity with $\mathsf{H}_{\rm MSC}$ by $0.4828$ bits/s in Melbourne and by a similar $0.4837$ bits/s in Helsinki. An important observation from these numerical results is that the difference between $\mathsf{H}_{\rm MSC}$ and $\mathsf{H}_{\rm MSC_0}$ remains unnoticeable for serving times up to $10$–$20$ seconds, characterising the sensitivity of the handover strategy to persistence. For longer serving times, $\mathsf{H}_{\rm MSC}$ outperforms both $\mathsf{H}_{\rm MSC_0}$ and $\mathsf{H}_{\rm Rand}$ by a growing margin, since they do not use any information regarding the orbit trajectory over the serving time; when the serving time is short, there is relatively little information to use for handover, however, as the serving time increases, there is more and more information that must be used to make the most informed decision possible---this is why long serving times result in more complicated handover strategies compared to those for short serving times.

The capacity upper bound $\overline{C}_{\rm pers}$ is significantly looser with fixed serving times compared to the earlier case with cap serving times. This may {be} justified through their respective heat maps in Fig. \ref{fig:ServCap}(c) and Fig. \ref{fig:ServCap}(d), which {show} significantly smaller light grey regions, reducing the probability of a visible satellite with a capacity near the upper bound. Nonetheless, the higher latitude user has a tighter bound, and in this scenario the bound is significantly tighter than {at lower latitudes}.

\section{Conclusion}

This paper introduced persistent capacity for a semi-stochastic mega-constellation channel model to 
{study} optimal handover strategies 
{under} general serving times, extending prior stochastic models that 
{are accurate primarily for short serving intervals}. Persistent capacity 
{provides an information-theoretic characterisation of} LEO satellite networks for a fixed ground user, and the handover strategy that maximises it was derived using a Dinkelbach-type algorithm for 
{nonlinear} fractional programs. 
{The resulting} tractable closed-form decision rule induces an ordering over preferred satellites for handover, which 
{may be} exploited in multi-user handover optimisation algorithms to reduce the search space and simplify large-scale assignment problems.

Numerical results showed that non-persistent capacity 
{closely approximates} persistent capacity 
{for serving times} up to around $15$ seconds, consistent with current Starlink operations. 
{For longer serving times, however, handover strategies that account for persistence become important} to avoid significant capacity degradation. Tight information-theoretic upper and lower bounds were 
{also} derived for 
{analytically intractable cases}, with the upper bound 
{being} particularly accurate for high-latitude users.

Overall, this work establishes persistent capacity as a tractable information-theoretic benchmark for LEO mega-constellation handover. 
{The framework provides both theoretical insight into the limits of user--satellite association and practical decision rules for designing handover strategies in large-scale satellite networks.}

\appendix
{
\subsection{Proof of Theorem \ref{thm:circ_orbit}}
Define the rotation matrices $R_x,R_y,R_z$ for rotations around $x,y,z$, respectively. Start with the satellite on the flat orbital plane (the $x$-$y$ plane) at $r_0(t) = [R\cos(\psi(t)), R\sin(\psi(t)),0]^{\top}$. Then we want to compute $r(t) = R_z(\theta(0))R_y(\phi(0) - \frac{\pi}{2})R_x(\beta(0))r_0(t)$. For the first two rotations, we have $R_y(\phi(0) - \frac{\pi}{2})R_x(\beta(0))r_0(t) = R[A(t),B(t),C(t)]^\top$. Since $R_z$ only rotates the $x$-$y$ components and leaves $z$ unchanged, we can apply the third rotation to get
    \begin{align}
        r(t) = 
        R\begin{bmatrix}
            A(t) \cos\theta(0) - B(t) \sin \theta(0) \\ A(t) \sin\theta(0) + B(t) \cos\theta(0) \\ C(t)
        \end{bmatrix}.\notag
    \end{align}
Now we convert from cartesian coordinates to spherical coordinates. Since $z(t) = RC(t)$, we have $\phi(t) = \cos^{-1}(z(t)/R) =\cos^{-1}(C(t))$. Longitude is the argument of the complex number $x(t)+jy(t)= e^{j \theta(0)} (A(t) + jB(t))$.
}

 \subsection{Closed-Form Persistent Capacity for Random Handover}\label{appendix:C_rand_HO}

Let $\Delta t \rightarrow 0$ so that the persistent capacity with random handover can be written as\footnote{Note that this is a tight approximation for any finite $\Delta t $ within the coherence time of the satellite path loss.} \cite{Mcbain2025b}
\begin{align}
    C_{\rm pers}[\mathsf{H}_{\rm Rand}] = \frac{\mathbb{E}\left[\widetilde{\mathsf{C}}(\Theta,\Phi,A)\right]}{\mathbb{E}[\mathsf{T}(\Theta,\Phi,A)]}
\end{align}
where 
\begin{align}
    &\widetilde{\mathsf{C}}(\Theta,\Phi,A)\notag\\
    &= \int_{0}^{\mathsf{T}(\Theta,\Phi,A)} \mathbb{E}_{|h(t)|^2}\left[\log \left(1 + \frac{\gamma |h(t)|^2}{d^2(\Theta(t),\Phi(t))}\right) \right] dt
\end{align}
for shadowed-Rician process $h(t)$, which is readily computable using numerical integration.  

The integration for computing the expectation $\mathbb{E}[f(\Theta,\Phi,A)]$ for each function $f \in \{\widetilde{\mathsf{C}}, \mathsf{T}\}$ over the cap region $\mathsf{Cap}$ is \cite{Mcbain2025b}
\begin{align}
   &\mathbb{E}[f(\Theta,\Phi,A)]\notag\\
   &= 
\frac{1}{4\pi}\int_{0}^{\pi} \int_{\theta_L(\phi)}^{\theta_U(\phi)} [f(\theta, \phi, -1)+f(\theta, \phi, 1)] f_{\Phi}(\phi)  d\theta d\phi .
\end{align}
Therefore, the persistent capacity with random handover can be computed using numerical integration.

\subsection{Properties of $Q_N(C)$}\label{appendix:Q_properties}

\begin{lemma}\label{lem:QN_convex}
    $Q_N(C)$ is convex.
\end{lemma}
\begin{proof} For $t\in[0,1]$ and $C,C' \in \mathbb{R}$, we have
\begin{equation}
\begin{aligned}
Q_N(tC + (1-t)C') 
&= \frac{1}{N} \sum_{n=1}^N \max_{1 \leq k \leq |\mathcal{V}|} \big\{ t (C_{k,n} - CN_{k,n}) \\
&\quad + (1-t) (C_{k,n} - C' N_{k,n}) \big\} \\
&\leq \frac{t}{N} \sum_{n=1}^N \max_{1 \leq k \leq |\mathcal{V}|} \{ C_{k,n} - CN_{k,n} \} \\
&\quad + \frac{1 - t}{N} \sum_{n=1}^N \max_{1 \leq k \leq |\mathcal{V}|} \{ C_{k,n} - C' N_{k,n} \} \\
&= t Q_N(C) + (1 - t) Q_N(C')
\end{aligned}
\label{eq:convexity}
\end{equation}
\end{proof}

\begin{lemma}\label{lem:QN_monotone}
    $Q_N(C) < Q_N(C')$ if $C > C'$.
\end{lemma}
\begin{proof} For $C,C' \in \mathbb{R}$ with $C > C'$, then for some $k(n)$ we have
    \begin{align}
        Q_N(C) &= \frac{1}{N}\sum_{n=1}^{N}  (C_{k(n),n} -CN_{k,n})\\
        &< \frac{1}{N}\sum_{n=1}^{N}  (C_{k(n),n} -C'N_{k,n})\\
        &\leq \frac{1}{N}\sum_{n=1}^{N}  \max_{1 \leq k \leq |\mathcal{V}^{(n)}|} \{C_{k,n} -C'N_{k,n}\} \\&= Q_N(C')\notag .
    \end{align}
\end{proof}

\begin{lemma}\label{lem:QN_zero}
    $Q_N(C^*)=0$ has a unique solution $C^*$.
\end{lemma}
\begin{proof}
Since $N_{k,n}>0$, $Q_N(C)\to-\infty$ as $C\to\infty$ and $Q_N(C)\to+\infty$ as $C\to-\infty$; combined with continuity, the Intermediate Value Theorem gives existence of $C^*$, and Lemma \ref{lem:QN_monotone} gives uniqueness.
\end{proof}

\bibliography{refs}
\bibliographystyle{IEEEtran}

\end{document}